\documentclass[nolinenumbers]{aastex631}

\usepackage{amsmath}

\accepted{Nov. 13, 2023}

\shorttitle{Strong resemblance between surface and deep zonal winds inside Jupiter}
\shortauthors{Cao et al.}

\graphicspath{{./}{figures/}}

\begin{document}

\title{Strong resemblance between surface and deep zonal winds inside Jupiter \\ revealed by high-degree gravity moments}

\correspondingauthor{Hao Cao}
\email{hcao@epss.ucla.edu}

\author[0000-0002-6917-8363]{Hao Cao}
\affiliation{Department of Earth, Planetary, and Space Sciences, \\ University of California, Los Angeles (UCLA), Los Angeles, CA 90095, USA}
\affiliation{Department of Earth and Planetary Sciences, \\
Harvard University, Cambridge, MA 02138, USA}

\author[0000-0001-8556-2675]{Jeremy Bloxham}
\affiliation{Department of Earth and Planetary Sciences, \\ Harvard University, Cambridge, MA 02138, USA}

\author[0000-0001-9896-4585]{Ryan S. Park}
\affiliation{Jet Propulsion Laboratory, California Institute of Technology, \\ Pasadena, CA 91011, USA}

\author[0000-0002-7092-5629]{Burkhard Militzer}
\affiliation{Department of Earth and Planetary Science, \\ University of California, Berkeley, CA 94720, USA}
\affiliation{Department of Astronomy, University of California, Berkeley, CA 94720, USA}

\author[0000-0002-9569-2438]{Rakesh K. Yadav}
\affiliation{Department of Earth and Planetary Sciences, \\
Harvard University, Cambridge, MA 02138, USA}

\author[0000-0001-8550-749X]{Laura Kulowski}
\affiliation{Department of Earth and Planetary Sciences, \\
Harvard University, Cambridge, MA 02138, USA}

\author[0000-0001-9432-7159]{David J. Stevenson}
\affiliation{Division of Geological and Planetary Sciences, \\
California Institute of Technology, Pasadena, CA 91125, USA}

\author[0000-0002-9115-0789]{Scott J. Bolton}
\affiliation{Southwest Research Institute, San Antonio, TX 78238, USA}

\begin{abstract}

Jupiter's atmosphere-interior is a coupled fluid dynamical system strongly influenced by the rapid background rotation. While the visible atmosphere features east-west zonal winds on the order of $\sim$100 m s$^{-1}$ \citep{Tollefson2017}, zonal flows in the dynamo region are significantly slower, on the order of $\sim$ cm s$^{-1}$ or less, according to the latest magnetic secular variation analysis \citep{Bloxham2022}. The vertical profile of the zonal flows and the underlying mechanism remain elusive. The latest Juno radio tracking measurements afforded the derivation of Jupiter's gravity field to spherical harmonic degree 40. Here, we use the latest gravity solution to reconstruct Jupiter's deep zonal winds without \textit{a priori} assumptions about their latitudinal profile. The pattern of our reconstructed deep zonal winds strongly resemble that of the surface wind within $\pm 35^\circ$ latitude from the equator, in particular the northern off-equatorial jet (NOEJ) and the southern off-equatorial jet (SOEJ) \citep{Kulowski2021}. The reconstruction features larger uncertainties in the southern hemisphere due to the north-south asymmetric nature of Juno’s trajectory. Amplitude of the reconstructed deep NOEJ matches that of the surface wind when the wind is truncated at a depth $\sim$2500 km, and becomes twice that of the surface wind if the truncation depth is reduced to $\sim$1500 km. Our analysis supports the physical picture in which prominent part of the surface zonal winds extends into Jupiter's interior significantly deeper than the water cloud layer.

\end{abstract}


\section{Introduction} \label{sec:intro}

Jupiter's visual appearance is dominated by belts, zones, and vortices of varying sizes. Shear in the east-west zonal flows and the associated upwellings and downwellings are closely linked to the belts and zones, although the origin of the alternating bands of zonal winds is still a topic of debate. Moist convection in the shallow atmosphere \citep{VS2005, schneider2009, liu2010}, deep convection in the molecular envelope \citep{Busse1976,heimpel2005,GastineWicht2021}, and tides from the moons of Jupiter \citep{Lindzen1991,Tyler2022} have been proposed as possible underlying mechanisms. Numerical models for each of these proposals have been constructed, with varying degrees of success in reproducing the observed zonal winds patterns and amplitude. 

Determination of the depth of the rapid atmospheric zonal flows provides a critical observational clue regarding their origin and dynamical balance, even though this depth would not directly translate to the driving mechanism \citep[e.g.,][]{Showman2006, Christensen2020}. Measurements of the gravitational \citep{hubbard1999, iess2018, kaspi2018} and magnetic field \citep{Cao2017b, moore2019, Cao2020, Bloxham2022} can be used to infer structure and dynamics inside giant planets. Gravitational harmonics are increasingly more sensitive to the outer layers as the spherical harmonic (SH) degree increases \citep{Guillot2004}. The magnetic field is sensitive to depth with sufficient electrical conductivity \citep{liu2008, Cao2017b}, e.g., greater than $\sim$2000 km inside Jupiter \citep{french2012}. 

Previous gravity solutions from the Juno radio tracking experiments \citep{Folkner2017, iess2018, Durante2020} have provided the zonal gravitational harmonics of Jupiter up to SH degree $\sim$10. These gravity solutions were employed to constrain the background structure \citep{dc2019, Miguel2022, Militzer2022} as well as dynamics, in particular the deep zonal wind structure, inside Jupiter \citep{kaspi2018, kong2018, Kulowski2021}. One continuing debate regarding the deep zonal winds inside Jupiter is to what extent they resemble the observed surface winds, as deep winds that are either quite similar to \citep{kaspi2018, kong2018, Kulowski2021, Militzer2022} or very different from \citep{kong2018} the surface winds have been constructed to match the gravity field up to SH degree $\sim$10. 

\subsection{Jupiter's surface winds in spectral space} \label{subsec:wind_in_spectral}

\begin{figure}[h]%
\centering
\includegraphics[width=0.9\textwidth]{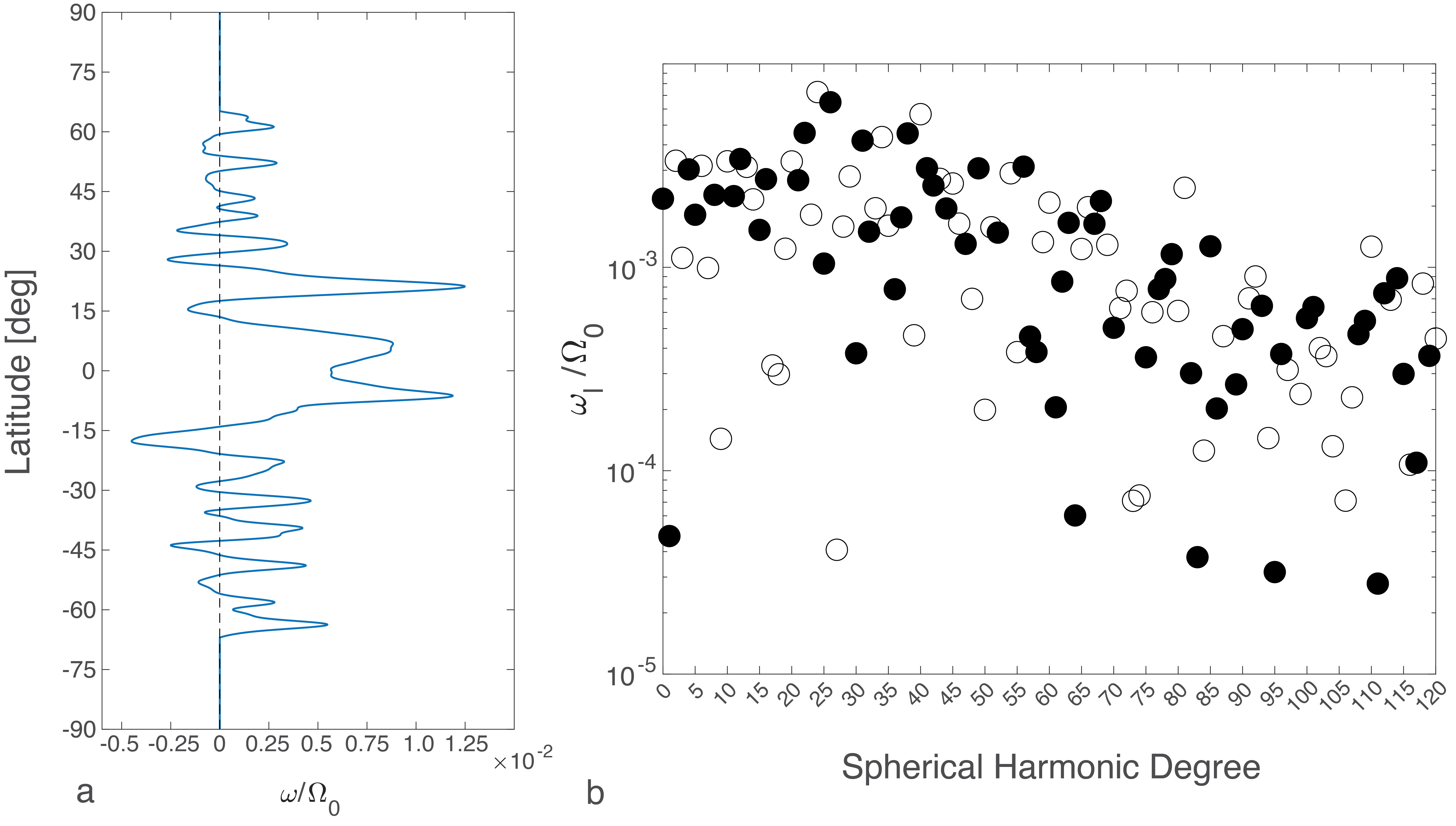}
\caption{Angular velocity of Jupiter's surface zonal winds in real space \citep{Tollefson2017} as a function of planetocentric latitude (a) and in spherical harmonic space (b). Here we show the ratio of the zonal winds angular velocity in the System III rotating frame, $\omega$, and the angular velocity of the background rotation $\Omega_0$. In Panel b, filled circles represent positive values while open circles represent negative values. It can be seen that the peak zonal wind speed is $\sim$1\% of the background rotation and the dominant latitudinal length-scale in the surface zonal wind corresponds to spherical harmonic degree $\sim$24.}\label{fig1}
\end{figure}

Here we point out that an important part of this debate stems from the limited spatial resolution of the previous gravity solutions. The typical latitudinal length-scale of the observed surface winds corresponds to SH degrees much higher than 10 (Fig. \ref{fig1}). As a result, slower and latitudinally smoother deep winds can be constructed if one only tries to match the large-scale gravity field with SH degree $n \le 10$. 

To illustrate this point, like the way we decompose the gravitational field onto a set of spherical harmonics, here we decompose the zonal wind angular velocity (in the rotating frame) onto the Legendre polynomials 
\begin{equation}
\omega(\theta)=\sum_{l=0} \omega_l P_l(\cos \theta),
\end{equation}
where $l$ is the SH degree, $\theta$ is the co-latitude measured from the north-pole, $P_l(\cos \theta)$ is the Legendre polynomial of degree $l$, and $\omega_l$ is the coefficient. Any residual solid-body rotation component is described with the degree-0 coefficient $\omega_0$. Even degree modes correspond to north-south symmetric winds, while odd degree modes correspond to north-south antisymmetric winds. 

The zonal wind velocity $U_\phi$ in the rotating frame is connected to the angular velocity via
\begin{align}
U_\phi(r, \theta) & =\omega(r, \theta) s, \\
                  & =\omega(r, \theta) r \sin \theta,
\end{align}
where $r$ is the spherical radial distance from the center of mass, and $s=r \sin \theta$ is the cylindrical radial distance to the spin-axis.

Fig. \ref{fig1} shows Jupiter's average surface zonal wind angular velocity \citep{Tollefson2017} in the System III rotating frame as a function of planetocentric latitude (panel a) as well as its decomposition onto the Legendre polynomials (panel b). It can be seen from Fig. \ref{fig1}b that the spectral content of the surface zonal winds peak around SH degree 24. Gravitational harmonics up to a similar wave-number would be needed to determine whether Jupiter's deep zonal wind pattern resembles that of the surface wind. Throughout this paper, we use $l$ to refer to the SH degree of the zonal wind spectral decomposition and $n$ to refer to the SH degree of the gravity harmonics. 

The degree-0 coefficient $\omega_0$ is non-zero for Jupiter's surface zonal wind when viewed in the System III rotating frame. This corresponds to a non-zero positive solid-body rotation component of Jupiter's surface zonal winds in this rotating frame defined by Jupiter's internal magnetic field. One can also see from Fig. \ref{fig1}a that the surface zonal wind viewed in this frame has a net positive (eastward) component, e.g., the eastward winds are stronger and there are more eastward winds in particular at low latitudes.

\subsection{An updated gravity field for Jupiter}\label{subsec:Jupiter_gravity_field}

\begin{figure}[h]%
\centering
\includegraphics[width=0.9\textwidth]{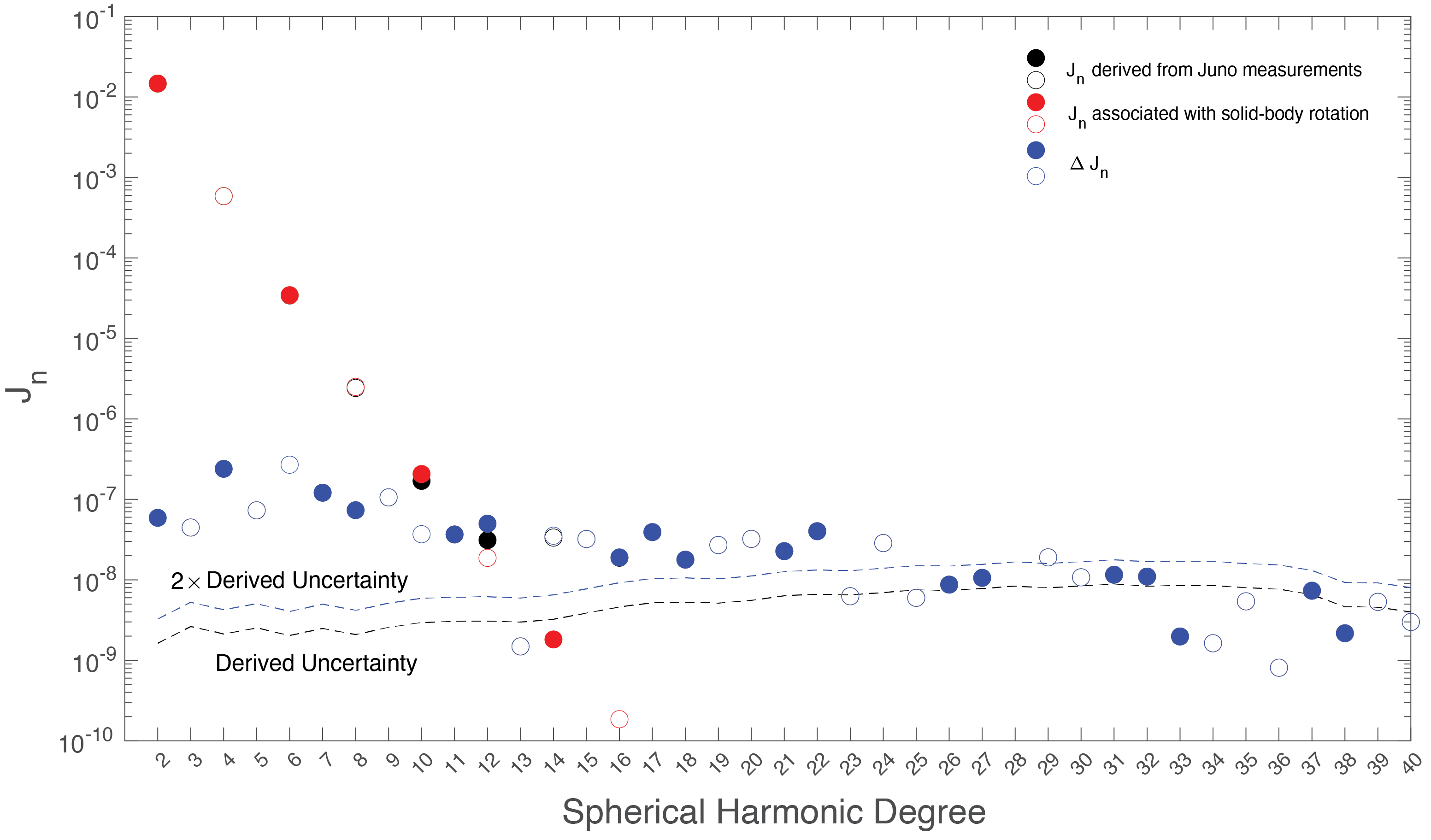}
\caption{The zonal harmonics of the latest gravity solution for Jupiter (black circles), the derived uncertainties (dashed lines) \citep{kaspi2023}, the solid-body rotation contribution (red circles) according to one of the latest interior models of Jupiter \citep{Militzer2022}, and the differences between the two (blue circles). For the $J_n$s, filled (open) circles represent positive (negative) values. Twice the derived uncertainties is about 3 times the formal uncertainty.}\label{fig2}
\end{figure}

A new gravity field solution for Jupiter \citep{kaspi2023} was recently derived by analyzing the Juno Doppler tracking data up to orbit PJ 37 (Fig. \ref{fig2}). In this new gravity solution, the small-scale gravitational accelerations beyond SH degree 12 were constrained to be zero near the poles \cite[e.g.,][]{KONOPLIV2020,Park2020}. More specifically, for latitudes between (90$^\circ$S, 40$^\circ$S) and (70$^\circ$N, 90$^\circ$N), we created a grid point for every two degrees in latitude. Then, for each grid point, we assumed the \textit{a priori} surface acceleration values due to zonal harmonics $J_{13}$ to $J_{40}$ are zero. The \textit{a priori} uncertainties for these surface acceleration points are empirically determined so that the mapped surface acceleration reaches the uncertainty 1 mGal (mili-Gal, where 1 Gal=1 $cm/s^2$). 

Fig. \ref{fig2} shows the zonal harmonics of this new gravity solution, $J_n$, in black circles and the derived uncertainties in black dashed line up to SH degree 40. The filled (open) circles represent positive (negative) values. The derived uncertainties shown here are the square root of the diagonal terms of the full covariance matrix. The derived uncertainties considered non-explicit factors in the formal inversion process, and are about 1.5 times the formal uncertainty ($1\sigma$). Thus, twice the derived uncertainty corresponds to about three times the formal uncertainty ($3\sigma$). It can be seen from Fig. \ref{fig2} that this new gravity solution resolves Jupiter’s $J_{n}$ up to SH degree 32 above the derived uncertainty, and up to SH degree 24 above twice the derived uncertainty. This new solution now resolves the length-scales in Jupiter’s gravity field that correspond to the typical (latitudinal) length-scales in Jupiter’s surface zonal flows ($\sim$ SH degree 24). Furthermore, this new high-resolution gravity solution affords an independent reconstruction/inversion of Jupiter's deep zonal winds without any input about the surface winds, which is the philosophy we adopt in this study. Our philosophy is in marked contrast to previous studies which project the observed surface winds into Jupiter's interior \citep[e.g.,][]{kaspi2018, kaspi2023}. 

Contributions from the solid-body-rotation (SBR) dominate the even degree $J_{n}$ of Jupiter up to SH degree $\sim$10 but become insignificant shortly after, as shown by the red circles in Fig. \ref{fig2} which correspond to the latest Jupiter interior model by \citet{Militzer2022}. Most of the even degree measured $J_{n}$ with $n \ge 12$ and all odd degrees $J_{n}$ must have a dynamical origin \citep{hubbard1999, kaspi2013}. They correspond to axisymmetric density/shape perturbations which are most likely in balance with deep zonal flows. It shall be noted that the details of this density-flow balance remain somewhat uncertain, in particular for the part where the wind decays with depth \cite[e.g., see discussion in][]{Kulowski2021}. At present, the only dynamical model for deep wind truncation inside Jupiter \citep{Christensen2020} asserts that the zonal winds are $z$-invariant from the near surface until a truncation depth where the wind amplitude decays sharply within a vertical range of 150 -- 300 km via a thermal wind shear mechanism. Here, the $z$-axis is aligned with the spin-axis of Jupiter. The latitudinal density gradient in this model \citep{Christensen2020} results from meridional circulation acting on the background non-adiabatic density gradient in a hypothesized stably stratified layer (SSL). Maxwell stress in the semi-conducting region could provide the driving force of the meridional circulation \citep{Christensen2020}. Reynolds stress is likely another option. 

\section{Methods and Assumptions} \label{sec:Methods}

Even with $\sim$30 gravitational harmonics, inferring the deep wind structure inside Jupiter is still a non-unique problem given the integral nature of the gravitational harmonics. \textit{A priori} assumptions about the spatial profile of the wind (latitudinal and/or vertical) and the force/vorticity balance are necessary ingredients of the solution, regardless of whether they are made explicit or not. Here we adopt the \citet{Christensen2020} deep zonal wind decay mechanism, but do not require the latitudinal profile of the deep zonal winds to be that of the observed surface wind. In fact, no information on the observed surface zonal winds was fed into our inversion analysis. In addition, we allow the truncation depth of the deep zonal winds to be a free parameter and scan the range between 1000 km and 3000 km. By adopting this philosophy, we could utilize the latest Jupiter gravity solution to clarify the resemblance between the surface and deep zonal winds above the highly conducting dynamo region \citep{Bloxham2022}. 

\subsection{Construction of an ensemble of deep zonal wind modes}

\begin{figure}[h]%
\centering
\includegraphics[width=0.9\textwidth]{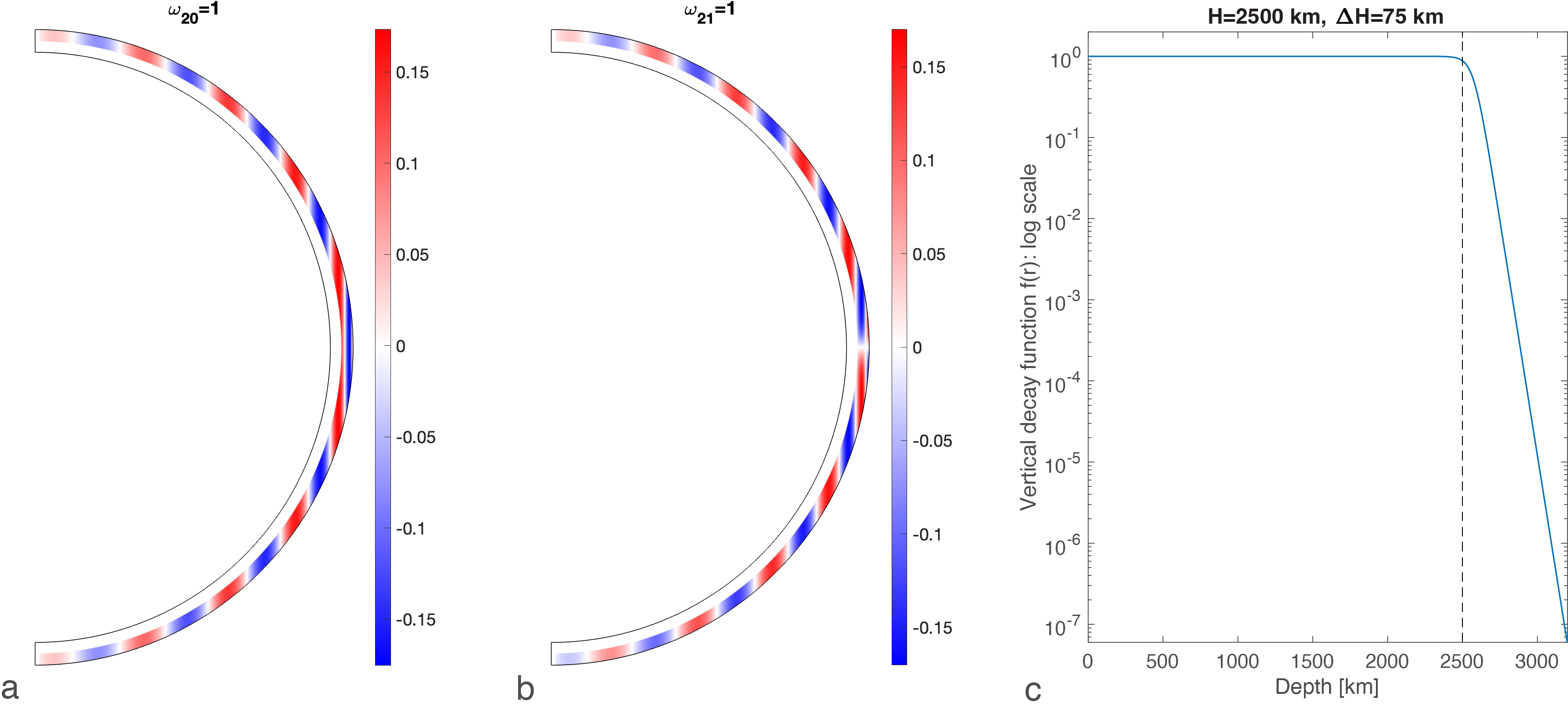}
\caption{Meridional structure of single mode deep zonal winds and the vertical truncation function. Panel a shows an equatorially symmetric (even) mode with $l=20$, panel b shows an equatorially antisymmetric (odd) mode with $l=21$. For both the even and the odd modes, a vertical truncation function (Eq. \ref{eq:vT}) with $H=2500$ km, and $\Delta H=75$ km has been applied (panel c). In addition, an equatorial smoothing function (Eq. \ref{eqn:equatorial_smoothing}) with $\Delta z=0.075$ $R_J$ has been applied to the odd mode (see Fig. \ref{fig:equatorial_smoothing}).}\label{fig:single_mode_radial_decay}
\end{figure}

We first construct an ensemble of deep zonal wind modes with Legendre polynomials in the latitudinal direction (Eq. \ref{eqn:single_mode_surface}) and hyperbolic tangent functions in the vertical direction (Eq. \ref{eq:vT}) and then forward compute the $J_n$ associated with these wind modes with the thermal wind balance (see section \ref{subsec:formulate_inv} for more details regarding the thermal wind balance adopted in this study). Guided by the \citet{Christensen2020} model, the thickness of the (vertical) transition layer is kept in the range of 150 -- 300 km. Fig. \ref{fig:single_mode_radial_decay}ab showcase two examples of individual deep zonal wind mode with truncation depth of 2500 km, corresponding to SH degree 20 and 21, respectively. 

The construction of an individual deep zonal wind mode starts with a surface wind angular velocity prescribed by a single Legendre polynomial
\begin{equation}
\omega(\theta_S)=\omega_l P_l(\cos \theta_S),
\label{eqn:single_mode_surface}
\end{equation}
where $\theta_S$ is the co-latitude at the surface of the planet and $\omega_l$ is the coefficient. For the north-south symmetric wind modes, which are  even degree modes, we first extend the surface winds along the spin-axis direction ($z$) invariantly and then apply the following hyperbolic vertical truncation function
\begin{equation}
f(r)=\frac{1}{2}\frac{\tanh \left( \frac{r+H-r_J}{\Delta H} +1 \right) +1}{\tanh \left( \frac{H}{\Delta_H} + 2 \right)}, \label{eq:vT}
\end{equation}
where $\tanh$ is the hyperbolic tangent function, $r_J$ is the radius of Jupiter, $H$ is the truncation depth, and $\Delta H$ is the parameter setting the thickness of the transition layer. $\Delta H$ is set to values between 50 km and 100 km for the cases presented in this study to achieve a transition thickness between 150 and 300 km as suggested by the theoretical model of \cite{Christensen2020}. Fig. \ref{fig:single_mode_radial_decay}c shows an example of the vertical truncation function with $H=2500$ km and $\Delta H=75$ km. 

\begin{figure}[h]%
\centering
\includegraphics[width=0.5\textwidth]{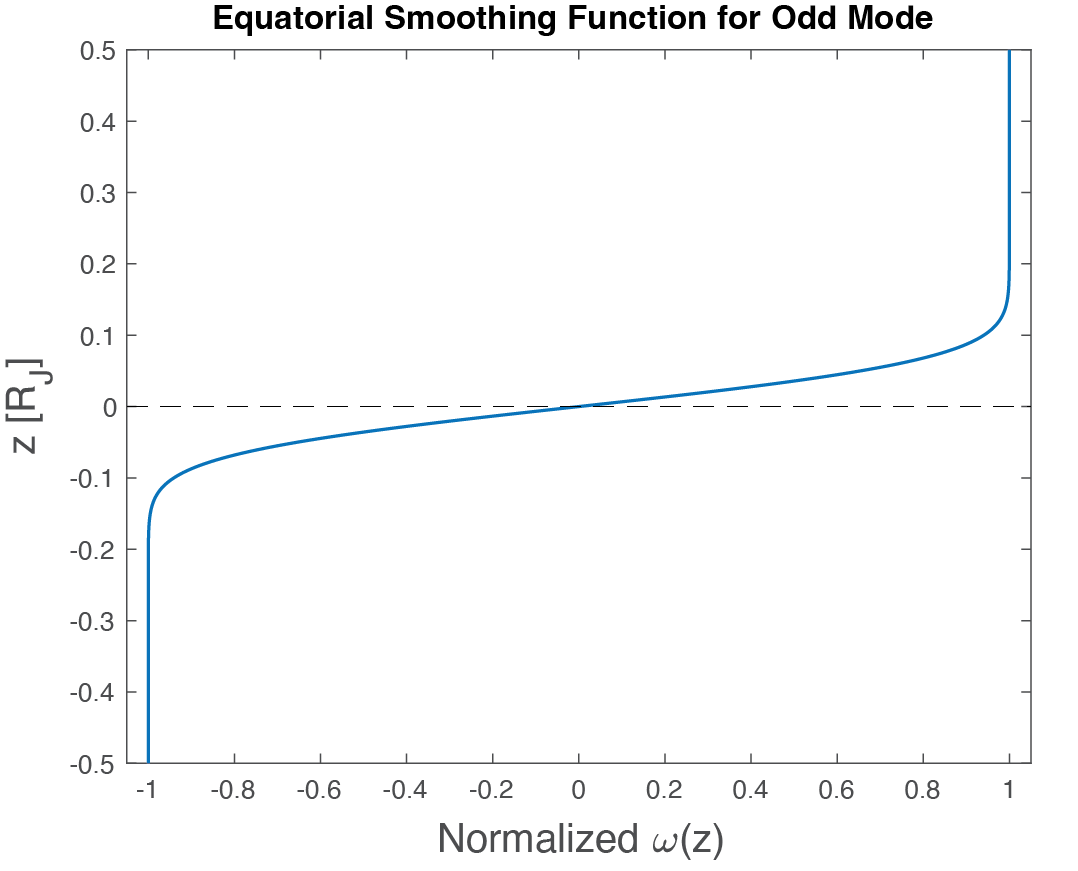}
\caption{Example of an equatorial smoothing function with $\Delta z=0.075$ $R_J$. An equatorial smoothing function (Eq. \ref{eqn:equatorial_smoothing}) is applied to the equatorial antisymmetric (odd) mode to ensure smooth transition between the nearly z-invariant winds in each hemisphere.}\label{fig:equatorial_smoothing}
\end{figure}

For north-south antisymmetric wind modes, which are the odd degree modes, we extend the surface winds along the spin-axis direction ($z$) invariantly in their respective hemisphere, then apply the following equatorial smoothing function, and further apply the vertical truncation function (Eq. \ref{eq:vT}),
\begin{equation}
\omega(s,z)=\omega_N (s_{surf}=s) \frac{1+erf \left( \frac{z}{\Delta z}\right)}{2} + \omega_S (s_{surf=s}) \frac{1- erf \left( \frac{z}{\Delta z}\right)}{2},
\label{eqn:equatorial_smoothing}
\end{equation}
where $\omega_N$ is the surface angular velocity in the northern hemisphere and $\omega_S$ is the surface angular velocity in the southern hemisphere. This ensures that the zonal wind in each hemisphere is nearly $z$-invariant except within a thin-layer near the equator. The influence of the choice of $\Delta z$ on the values of odd $J_n$ up to SH degree 40 associated with each invidual mode is minimal for all $\Delta z < 0.08 r_J$. In all the cases presented in this study, $\Delta z$ is set to 0.075 $r_J$. Moreover, the reconstructed deep winds are found to be mostly north-south symmetric within $\pm$15$^\circ$ from the equator, which implies that solutions with a truncation depth $\le$ 2500 km are essentially unaffected by the equatorial smoothing function. Fig. \ref{fig:equatorial_smoothing} shows an example of the equatorial smoothing function corresponding to 30$^\circ$ surface latitude. Here $\omega_N$ is normalized to 1 while $\omega_S$ is normalized to -1. 

\subsection{Formulate the deep wind-gravity field connection as a linear inverse problem} \label{subsec:formulate_inv}

For each deep zonal wind mode, referred to as $\omega_l$ for simplicity, we can forward compute the associated gravitational harmonics $J_n^{\omega_l}$ for all $n$ of interest, which are dominated by the ones with the same symmetry properties around SH degree $l$. $n_{\max}$ is set to 40 as constrained by the available gravity solution, and we tested $l_{\max}$ between 40 and 60 and obtained broadly similar results. In all results presented in this paper, $l_{\max}$ is set to 50. In our forward computation, we adopt the thermal wind balance \citep{kaspi2013,Cao2017a,kaspi2018} with spherical background density $\rho_0(r)$ and internal gravitational acceleration $\mathbf{g_0}(r)$ provided by the latest Jupiter interior model \citep{Militzer2022}. More specifically, we adopt the following thermal wind balance for both the z-invariant and the z-varying part of the deep zonal wind
\begin{equation}
2|\Omega_0|\frac{\partial [ \rho_0(r) U_\phi \mathbf{e}_\phi]}{\partial z}=-\frac{1}{r}\frac{\partial \rho'}{\partial \theta} \mathbf{e}_\theta \times [-|g_0(r)| \mathbf{e}_r],
\label{eqn:TWE}
\end{equation}
in which $\rho'$ is the wind-induced density perturbation and $\mathbf{e}_r$, $\mathbf{e}_\theta$, and $\mathbf{e}_\phi$ are unit vectors in the spherical radial, co-latitudinal, and azimuthal directions, respectively. This choice is based on the theoretical work of \cite{Christensen2020} and supported by the 3D numerical modeling of \cite{GastineWicht2021}, in which the $z$-varying part of the zonal winds is balanced by a latitudinal gradient in the non-adiabatic density (thermal wind shear). This is different from the barotropic zonal wind model adopted by \cite{Kulowski2021}, in which Eq. (\ref{eqn:TWE}) is only applied to the $z$-invariant part of the zonal winds.

With a given deep zonal wind profile, $U_\phi(r,\theta)$, one can integrate Eq. (\ref{eqn:TWE}) in the $\theta$ direction to get the wind-induced density perturbation, $\rho'(r,\theta)$, and then compute the wind-induced gravity harmonics via the volume integral
\begin{equation}
J_n=-\frac{1}{M a^n} \int_V \rho'(\mathbf{r}) r^n P_n (\cos \theta)dV,
\end{equation}
where $M$ is the total mass of Jupiter, $a$ is the 1-bar equatorial radius of Jupiter, and $V$ is the entire volume of Jupiter. Interested readers can refer to \cite{Cao2017a} for the derivation of eqn. (\ref{eqn:TWE}) and related technical details.

For each truncation depth, we can then formulate the deep wind-gravity field connection as a linear inverse problem. Formally, one could write the forward problem as 
\begin{equation}
\mathbf{obs} = G \, \mathbf{model}, 
\label{eqn:fwd_symbolic}
\end{equation}
here the observations, $\mathbf{obs}$, are the $\Delta J_n^{obs}$, which is the difference between the measured $J_n$ and those associated with solid-body-rotation $J_n^{SBR}$, the model parameters, $\mathbf{model}$, are the the coefficients $\omega_l$ associated with the deep zonal winds, while the forward matrix, $G$, consists of the (forward computed) wind-induced gravity harmonics, $J_n^{\omega_l}$, associated with the individual zonal wind mode $\omega_l$. To be more specific, one can write the above forward problem as 
\begin{equation}
\begin{bmatrix}
\Delta J_2 \\
\Delta J_3 \\
\Delta J_4 \\
\vdots \\
\Delta J_n
\end{bmatrix}
=
\begin{bmatrix}
J_2^{\omega_0} & J_2^{\omega_1} & 
J_2^{\omega_2} & \dots & J_2^{\omega_l} \\

J_3^{\omega_0} & J_3^{\omega_1} & J_3^{\omega_2} & \dots & J_3^{\omega_l} \\

J_4^{\omega_0} & J_4^{\omega_1} & J_4^{\omega_2} & \dots & J_4^{\omega_l} \\

\vdots & \vdots & \vdots & \ddots & \vdots \\
J_n^{\omega_0} & J_n^{\omega_1} & J_n^{\omega_2} & \dots & J_n^{\omega_l}

\end{bmatrix}
\begin{bmatrix}
\omega_0 \\
\omega_1 \\
\omega_2 \\
\vdots \\
\omega_l
\end{bmatrix},
\label{eqn:fwd_specific}
\end{equation}
for which the inverse can be computed with standard linear inversion techniques such as the singular-value-decomposition (SVD). 

Formulating the deep wind-gravity field connection as a linear inverse problem also allows us to quantify the uncertainties of the solution based on the uncertainties in the observations. The error propagation starts with the full covariance matrix of the derived $J_n$ up to SH degree 40, which we denote as $data_{cov}$. The model covariance matrix, $mod_{cov}$, which describes the uncertainties and the correlations of the model parameters can be computed via 
\begin{equation}
mod_{cov}=G^{-1} \, data_{cov} \, \left(G^{-1}\right)^{T},
\label{eqn:mod_cov}
\end{equation}
where $G^{-1}$ is the inverse matrix and the superscript $T$ represents matrix transpose. Here, the model parameters are the coefficients $\omega_l$ associated with the reconstructed zonal winds. 

One can then compute the uncertainties associated with the solution in real space, $\Delta \omega(\theta)$, via 
\begin{equation}
\left[ \Delta \omega (\theta_S) \right]^2= \text{Fwd}(\theta_S) \, mod_{cov} \, \left[ \text{Fwd}(\theta_S) \right]^T, 
\label{eqn:delta_omega}
\end{equation}
where $\text{Fwd}$ is the forward matrix connecting $\omega(\theta)$ and $\omega_l$ evaluated at $\theta_S$
\begin{equation}
\text{Fwd}(\theta_S)=\left[P_0(\cos \theta_S), P_1(\cos \theta_S) , \dots , P_l(\cos \theta_S) \right].
\end{equation}

\section{Results} \label{sec:results}

With our methodology, we can solve for Jupiter's deep zonal winds at any truncation depth from a given set of $\Delta J_n^{obs}$ by inverting Eq. (\ref{eqn:fwd_specific}). We remove the solid-body rotation gravitational harmonics, $J_n^{SBR}$, corresponding to the \cite{Militzer2022} Jupiter model up to SH degree 16 from the observed $J_n$ of Jupiter \citep{kaspi2023} to get the $\Delta J_n^{obs}$. There are still some uncertainties in the SBR $J_n$ of Jupiter, in particular for $J_6$ \cite[cf.][]{Miguel2022}. The \cite{Militzer2022} Jupiter structural model with a dilute core requires a dynamical contribution to $J_6$ of $-0.27 \times 10^{-6}$. However, low-degree $\Delta J_n$, even or odd, correspond to long-wavelength features in the deep zonal flows and have relatively weak connection to the observed narrow zonal jets off the equator (see Fig. \ref{fig1}). 

\begin{figure}[h]%
\centering
\includegraphics[width=0.45\textwidth]{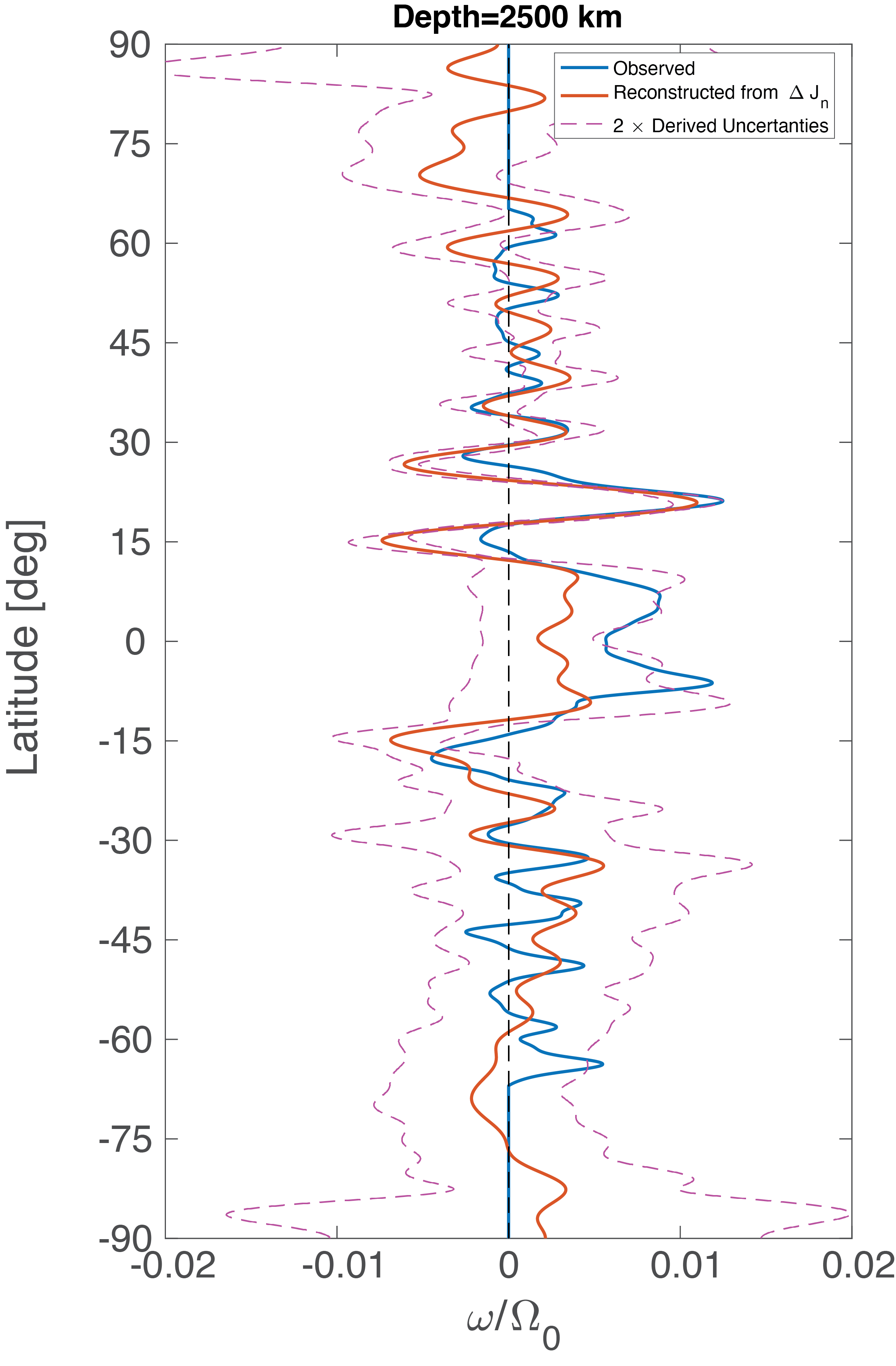}
\caption{Angular velocity of the reconstructed deep zonal winds with a truncation depth of 2500 km (red trace) compared to the observed surface zonal winds (blue trace) as function of surface planetocentric latitude. Also shown are the uncertainties associated with the reconstruction (dashed lines), computed from the full covariance matrix of the gravity solution. Strong resemblance between the observed surface zonal winds and reconstructed deep zonal winds, in both pattern and amplitude, are visible.}\label{fig3}
\end{figure}

Fig. \ref{fig3} compares the angular velocity of the reconstructed deep zonal winds with a truncation depth of 2500 km (solid red line) to the observed surface zonal winds (solid blue line) as a function of surface planetocentric latitude. Also shown are the uncertainties associated with this reconstruction (magenta dashed lines), computed/propagated from the full covariance matrix of the gravity solution following Eqs. (\ref{eqn:mod_cov}) \& (\ref{eqn:delta_omega}). The uncertainties here correspond to twice the derived uncertainties in $J_n$ ($\sim$3 times the formal uncertainties). It can be seen that there is a strong resemblance between the nominal reconstructed deep zonal winds and the observed surface winds, in particular the northern off-equatorial jet (NOEJ) around 20$^\circ$N and the southern off-equatorial jet (SOEJ) at a similar latitude in the southern hemisphere \citep{Kulowski2021}. For this particular truncation depth, the amplitude of the reconstructed prograde NOEJ also matches that of the observed NOEJ, which could indicate a local barotropic atmosphere at this location. The strong resemblance between the reconstructed deep wind (the nominal solution with $H=2500$ km) and the observed surface wind is mostly confined within $\pm 35^\circ$ latitude from the equator. Another feature of the reconstructed deep winds is the more pronounced north-south symmetry of the first retrograde jet adjacent to the prograde equatorial jet compared to the observed surface winds at Jupiter. This feature resembles the observed surface winds and inferred deep winds at Saturn \citep{GARCIAMELENDO2011, galanti2019, militzer2019}, as well as nearly all numerical deep convection models for Jupiter (and Saturn) \citep[e.g.,][]{heimpel2005, GastineWicht2021}. It is interesting to note that, while the pattern of the reconstructed mid-to-high latitude winds seems to deviate from that observed, the amplitude of the two remain very close to each other. Here we emphasize that no information about the observed surface winds was used in the inversion. Our reconstructed wind is an emergent solution from inverting Eq. (\ref{eqn:fwd_specific}) with the gravity harmonics of Jupiter and our forward model. 

Two aspects of the uncertainties associated with the reconstructed deep winds are worthy of comment. First, the uncertainties in the southern hemisphere are appreciably larger than that in the northern hemisphere. This results from the north-south asymmetric nature of Juno's orbits with increasingly more low-altitude coverage in the northern hemisphere due to the northward precession of Juno's periapsis. Second, the uncertainties within $\pm$12$^\circ$ of the equator are as large as those in the southern mid-latitudes. This is due to geometric effects associated with the shape of the planet and $z$-projection of the zonal winds, as 1) the $z$-derivative of the background density decreases towards zero as one approaches the lower latitude and 2) less mass is involved in a $z$-aligned column (annulus) with the same horizontal width at lower latitude. The observed prograde equatorial jet is within the uncertainties of the reconstructed deep zonal winds. Furthermore, since we are working with a spherical harmonic reconstruction of the deep winds and our knowledge of Jupiter's gravity harmonics (and their uncertainties) are currently limited to SH degree 40 (instead of, e.g., SH degree 100), these limit the computed uncertainties very close to the equator (e.g., within a few degrees from the equator). The true uncertainties very close to the equator is expected to be higher, as the geometric effect monotonically strengthens towards the equator. 

\begin{figure}[h]%
\centering
\includegraphics[width=0.45\textwidth]{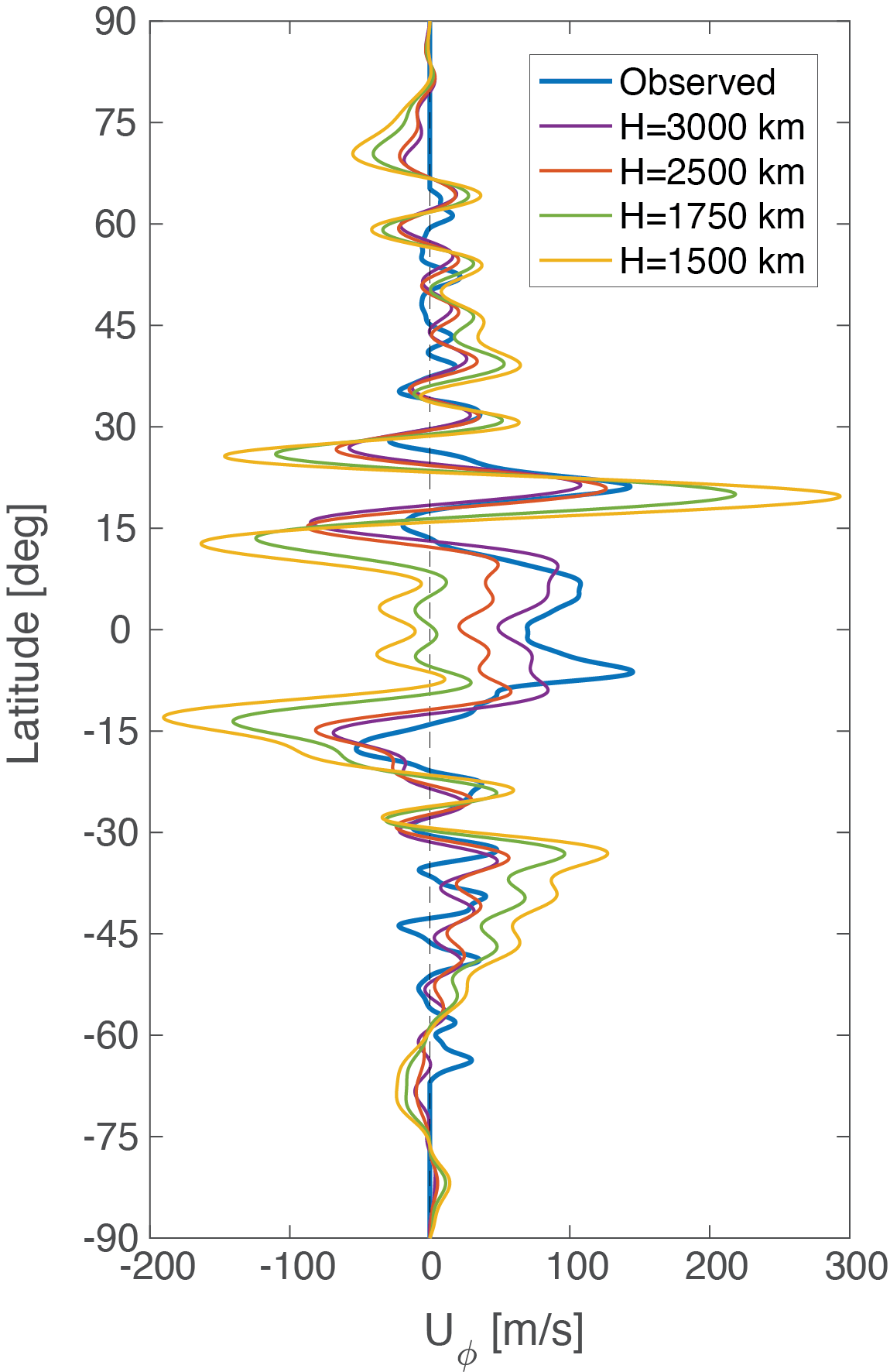}
\caption{Pattern and amplitude of the reconstructed deep zonal flows at a few different truncation depths. The observed surface zonal winds is also shown for comparison. It can be seen that the pattern of the reconstructed deep zonal flows remains broadly similar, while the amplitude increase rapidly as the truncation depth decrease.}\label{fig4}
\end{figure}

As our wind reconstruction procedure is not bounded by the observed surface winds, we could also investigate the effect of the truncation depth on the deep zonal flow amplitude-profile. Fig. \ref{fig4} displays an ensemble of reconstructed deep zonal wind as a function of the truncation depth. It can be seen that the latitudinal profile of the reconstructed deep zonal winds with different truncation depths remain broadly similar, in particular at latitudes northward of $\sim$30$^\circ$S and the prominent prograde NOEJ (keeping in mind that the reconstruction features higher uncertainties in the southern hemisphere). For truncation depth $>$ 2500 km, the amplitude of the reconstructed wind remains on the order of 100 $m/s$. The deeper most portion of these winds would violate the Ohmic dissipation constraint that results from the rapidly increasing electrical conductivity as a function of depth \citep{liu2008,Cao2017b}. The amplitude of the reconstructed deep winds increases as the truncation depth decreases, with faster rates at shallower truncation depth. The peak amplitude of the prograde NOEJ reaches $\sim$ 280 m s$^{-1}$, which is about twice that of the surface NOEJ, when the truncation depth is $\sim$ 1500 km. The peak amplitude reaches $\sim$ 780 m s$^{-1}$ (480 m s$^{-1}$) for a truncation depth of 1000 km (1250 km), due to the rapidly decreasing background density at shallower depth. For comparison, the sound speed at 1-bar is $\sim$ 1 km s$^{-1}$ \citep{Lorenz1998}, and increases to $\sim$ 4 km s$^{-1}$ at a depth $\sim$  1000 km \citep{french2012}. 

The Galileo probe, which was dropped into Jupiter around 6.5$^\circ$N and 4.4$^\circ$W (measured in the Jupiter System III coordinate) on 7th of December 1995 measured a factor of two increase in local zonal wind speed, from $\sim$ 85 m s$^{-1}$ at 1-bar to $\sim$ 170 m s$^{-1}$ at 5-bar, which then remain at the high speed level until about 21-bar \citep{Atkinson1997}. If similar near-surface shear also prevails at other latitudes (in particular near the NOEJ), our analysis indicates that the depth of such stronger winds would be shallower, $\sim$ 1500 km, though still significantly deeper than the expected water cloud layers. As discussed earlier, deep winds very close to the equator have relatively weak contributions to the gravity field due to geometrical effects. As a result, the true uncertainties of the derived deep winds very close to the equator, e.g. at the Galileo probe latitude, are expected to be large. Moreover, the uncertainties in the solid-body-rotation contribution to low-degree even $J_ns$ can affect the value of the derived equatorial jet as well as the pattern of the large-scale deep zonal wind in the latitudinal direction. 

\section{Summary and Discussion}\label{sec:discussion}

Adopting vertical profiles guided by the Christensen model \citep{Christensen2020, GastineWicht2021} but making no \textit{a priori} assumptions about the latitudinal profile, we constructed a suite of deep zonal winds inside Jupiter based on the latest gravity solution up to SH degree 40 \citep{kaspi2023}. The latitudinal profile of our reconstructed deep zonal flows bear a strong resemblance to the observed surface zonal winds within $\pm$35$^\circ$ latitude from the equator, in particular the prograde NOEJ around 20$^\circ$N. The NOEJ has been shown to dominate the low-degree odd gravity harmonics ($J_5$, $J_7$, $J_9$) even when assuming the vertical wind shear is balanced by Reynolds/Maxwell-stress instead of latitudinal entropy/composition shear (thermal wind balance) by \cite{Kulowski2021}. Moreover, our reconstruction demonstrated that latitudinally narrow zonal flows on the order of 100 m s$^{-1}$ or stronger reside in the top $\sim$ 2000 km of Jupiter. Our results thus strongly support the picture that the amplitude and pattern of the deep zonal winds in the non-electrically conducting part of the molecular envelope, which extends much deeper than the water cloud layer, are comparable to those observed at the surface. It remains a critical open question whether this picture could be reconciled with dynamical models without a stably stratified layer at the truncation depth. At present, no structure models require/invoke an SSL near this depth \citep[e.g.,][]{dc2019,Miguel2022,Militzer2022}. Answering this question has important implications not only for rotating fluid dynamics but also for the structure and evolution of Jupiter. A near-surface SSL can significantly impact the temperature profile in the deeper part of Jupiter as well as the cooling efficiency of the entire planet.  \\

\noindent \textbf{Acknowledgments} All authors acknowledge support from the NASA Juno project. Part of this research was carried out at the Jet Propulsion Laboratory, California Institute of Technology, under a contract with the National Aeronautics and Space Administration. H.C. would like to thank the Isaac Newton Institute for Mathematical Sciences, Cambridge, for support and hospitality during the programme \textit{Frontiers in dynamo theory: from the Earth to the stars} where some work on this paper was undertaken. This work was supported by EPSRC grant no EP/R014604/1. In addition, this work was partially supported by a grant from the Simons Foundation. \\

\noindent \textbf{Author contributions}
H.C. conceptualized the study, carried out the reconstruction of deep winds from gravity harmonics, and drafted the manuscript. J.B., R.K.Y., and L.K. supported the analysis and interpretation of the results. R.S.P. derived the gravity harmonics of Jupiter and their uncertainties from the Juno Radio Tracking experiments. B.M. provided the gravity harmonics associated with the solid-body rotating Jupiter. All authors contributed to the discussion, as well as editing and revising the manuscript. D.J.S. leads the Interior Working Group of Juno, and S.J.B. is the principal investigator of the Juno Mission. 


\noindent \textbf{Data Availability} All the Juno radio tracking data used to derive the gravity moments and their uncertainties in this study are available from the NASA Planetary Data System (https://pds.nasa.gov) at \url{https://doi.org/10.17189/1518938}.

\bibliography{JupiterWindsGravity_2023_HC_ApJ}{}

\begin{thebibliography}{}
\expandafter\ifx\csname natexlab\endcsname\relax\def\natexlab#1{#1}\fi
\providecommand{\url}[1]{\href{#1}{#1}}
\providecommand{\dodoi}[1]{doi:~\href{http://doi.org/#1}{\nolinkurl{#1}}}
\providecommand{\doeprint}[1]{\href{http://ascl.net/#1}{\nolinkurl{http://ascl.net/#1}}}
\providecommand{\doarXiv}[1]{\href{https://arxiv.org/abs/#1}{\nolinkurl{https://arxiv.org/abs/#1}}}

\bibitem[{{Atkinson} {et~al.}(1997){Atkinson}, {Ingersoll}, \& {Seiff}}]{Atkinson1997}
{Atkinson}, D.~H., {Ingersoll}, A.~P., \& {Seiff}, A. 1997, Nature, 388, 649, \dodoi{10.1038/41718}

\bibitem[{Bloxham {et~al.}(2022)Bloxham, Moore, Kulowski, Cao, Yadav, Stevenson, Connerney, \& Bolton}]{Bloxham2022}
Bloxham, J., Moore, K.~M., Kulowski, L., {et~al.} 2022, Journal of Geophysical Research: Planets, 127, e2021JE007138, \dodoi{10.1029/2021JE007138}

\bibitem[{{Busse}(1976)}]{Busse1976}
{Busse}, F.~H. 1976, Icarus, 29, 255, \dodoi{10.1016/0019-1035(76)90053-1}

\bibitem[{Cao {et~al.}(2020)Cao, Dougherty, Hunt, Provan, Cowley, Bunce, Kellock, \& Stevenson}]{Cao2020}
Cao, H., Dougherty, M.~K., Hunt, G.~J., {et~al.} 2020, Icarus, 344, 113541, \dodoi{10.1016/j.icarus.2019.113541}

\bibitem[{Cao \& Stevenson(2017{\natexlab{a}})}]{Cao2017b}
Cao, H., \& Stevenson, D.~J. 2017{\natexlab{a}}, Icarus, 296, 59, \dodoi{10.1016/j.icarus.2017.05.015}

\bibitem[{Cao \& Stevenson(2017{\natexlab{b}})}]{Cao2017a}
---. 2017{\natexlab{b}}, Journal of Geophysical Research: Planets, 122, 686, \dodoi{10.1002/2017JE005272}

\bibitem[{Christensen {et~al.}(2020)Christensen, Wicht, \& Dietrich}]{Christensen2020}
Christensen, U.~R., Wicht, J., \& Dietrich, W. 2020, The Astrophysical Journal, 890, 61, \dodoi{10.3847/1538-4357/ab698c}

\bibitem[{{Debras} \& {Chabrier}(2019)}]{dc2019}
{Debras}, F., \& {Chabrier}, G. 2019, The Astrophysical Journal, 872, 100, \dodoi{10.3847/1538-4357/aaff65}

\bibitem[{{Durante} {et~al.}(2020){Durante}, {Parisi}, {Serra}, {Zannoni}, {Notaro}, {Racioppa}, {Buccino}, {Lari}, {Gomez Casajus}, {Iess}, {Folkner}, {Tommei}, {Tortora}, \& {Bolton}}]{Durante2020}
{Durante}, D., {Parisi}, M., {Serra}, D., {et~al.} 2020, Geophysical Research Letters, 47, e86572, \dodoi{10.1029/2019GL086572}

\bibitem[{Folkner {et~al.}(2017)Folkner, Iess, Anderson, Asmar, Buccino, Durante, Feldman, Gomez~Casajus, Gregnanin, Milani, Parisi, Park, Serra, Tommei, Tortora, Zannoni, Bolton, Connerney, \& Levin}]{Folkner2017}
Folkner, W.~M., Iess, L., Anderson, J.~D., {et~al.} 2017, Geophysical Research Letters, 44, 4694, \dodoi{10.1002/2017GL073140}

\bibitem[{French {et~al.}(2012)French, Becker, Lorenzen, Nettelmann, Bethkenhagen, Wicht, \& Redmer}]{french2012}
French, M., Becker, A., Lorenzen, W., {et~al.} 2012, The Astrophysical Journal Supplement Series, 202, 5, \dodoi{10.1088/0067-0049/202/1/5}

\bibitem[{Galanti {et~al.}(2019)Galanti, Kaspi, Miguel, Guillot, Durante, Racioppa, \& Iess}]{galanti2019}
Galanti, E., Kaspi, Y., Miguel, Y., {et~al.} 2019, Geophysical Research Letters, 46, 616, \dodoi{10.1029/2018GL078087}

\bibitem[{García-Melendo {et~al.}(2011)García-Melendo, Pérez-Hoyos, Sánchez-Lavega, \& Hueso}]{GARCIAMELENDO2011}
García-Melendo, E., Pérez-Hoyos, S., Sánchez-Lavega, A., \& Hueso, R. 2011, Icarus, 215, 62, \dodoi{10.1016/j.icarus.2011.07.005}

\bibitem[{Gastine \& Wicht(2021)}]{GastineWicht2021}
Gastine, T., \& Wicht, J. 2021, Icarus, 368, 114514, \dodoi{10.1016/j.icarus.2021.114514}

\bibitem[{{Guillot} {et~al.}(2004){Guillot}, {Stevenson}, {Hubbard}, \& {Saumon}}]{Guillot2004}
{Guillot}, T., {Stevenson}, D.~J., {Hubbard}, W.~B., \& {Saumon}, D. 2004, {The interior of Jupiter}, ed. F.~{Bagenal}, T.~E. {Dowling}, \& W.~B. {McKinnon}, Vol.~1 (Cambridge University Press), 35--57

\bibitem[{Heimpel {et~al.}(2005)Heimpel, Aurnou, \& Wicht}]{heimpel2005}
Heimpel, M., Aurnou, J., \& Wicht, J. 2005, Nature, 438, 193, \dodoi{10.1038/nature04208}

\bibitem[{Hubbard(1999)}]{hubbard1999}
Hubbard, W.~B. 1999, Icarus, 137, 357, \dodoi{10.1006/icar.1998.6064}

\bibitem[{Iess {et~al.}(2018)Iess, Folkner, Durante, Parisi, Kaspi, Galanti, Guillot, Hubbard, Stevenson, Anderson, {et~al.}}]{iess2018}
Iess, L., Folkner, W., Durante, D., {et~al.} 2018, Nature, 555, 220, \dodoi{10.1038/nature25776}

\bibitem[{Kaspi(2013)}]{kaspi2013}
Kaspi, Y. 2013, Geophysical research letters, 40, 676, \dodoi{10.1029/2012GL053873}

\bibitem[{Kaspi {et~al.}(2018)Kaspi, Galanti, Hubbard, Stevenson, Bolton, Iess, Guillot, Bloxham, Connerney, Cao, {et~al.}}]{kaspi2018}
Kaspi, Y., Galanti, E., Hubbard, W.~B., {et~al.} 2018, Nature, 555, 223, \dodoi{10.1038/nature25793}

\bibitem[{{Kaspi} {et~al.}(2023){Kaspi}, {Galanti}, {Park}, {Duer}, {Gavriel}, {Durante}, {Iess}, {Parisi}, {Buccino}, {Guillot}, {Stevenson}, \& {Bolton}}]{kaspi2023}
{Kaspi}, Y., {Galanti}, E., {Park}, R.~S., {et~al.} 2023, Nature Astronomy, \dodoi{10.1038/s41550-023-02077-8}

\bibitem[{Kong {et~al.}(2018)Kong, Zhang, Schubert, \& Anderson}]{kong2018}
Kong, D., Zhang, K., Schubert, G., \& Anderson, J.~D. 2018, Proceedings of the National Academy of Sciences, 115, 8499, \dodoi{10.1073/pnas.1805927115}

\bibitem[{Konopliv {et~al.}(2020)Konopliv, Park, \& Ermakov}]{KONOPLIV2020}
Konopliv, A., Park, R., \& Ermakov, A. 2020, Icarus, 335, 113386, \dodoi{10.1016/j.icarus.2019.07.020}

\bibitem[{Kulowski {et~al.}(2021)Kulowski, Cao, Yadav, \& Bloxham}]{Kulowski2021}
Kulowski, L., Cao, H., Yadav, R.~K., \& Bloxham, J. 2021, Journal of Geophysical Research: Planets, 126, e2020JE006795, \dodoi{10.1029/2020JE006795}

\bibitem[{{Lindzen}(1991)}]{Lindzen1991}
{Lindzen}, R.~S. 1991, Geophysical and Astrophysical Fluid Dynamics, 58, 123, \dodoi{10.1080/03091929108227335}

\bibitem[{Liu {et~al.}(2008)Liu, Goldreich, \& Stevenson}]{liu2008}
Liu, J., Goldreich, P.~M., \& Stevenson, D.~J. 2008, Icarus, 196, 653, \dodoi{10.1016/j.icarus.2007.11.036}

\bibitem[{Liu \& Schneider(2010)}]{liu2010}
Liu, J., \& Schneider, T. 2010, Journal of the Atmospheric Sciences, 67, 3652, \dodoi{10.1175/2010JAS3492.1}

\bibitem[{{Lorenz}(1998)}]{Lorenz1998}
{Lorenz}, R.~D. 1998, Planetary and Space Science, 47, 67, \dodoi{10.1016/S0032-0633(98)00099-3}

\bibitem[{{Miguel} {et~al.}(2022){Miguel}, {Bazot}, {Guillot}, {Howard}, {Galanti}, {Kaspi}, {Hubbard}, {Militzer}, {Helled}, {Atreya}, {Connerney}, {Durante}, {Kulowski}, {Lunine}, {Stevenson}, \& {Bolton}}]{Miguel2022}
{Miguel}, Y., {Bazot}, M., {Guillot}, T., {et~al.} 2022, Astronomy \& Astrophysics, 662, A18, \dodoi{10.1051/0004-6361/202243207}

\bibitem[{Militzer {et~al.}(2019)Militzer, Wahl, \& Hubbard}]{militzer2019}
Militzer, B., Wahl, S., \& Hubbard, W. 2019, The Astrophysical Journal, 879, 78, \dodoi{10.3847/1538-4357/ab23f0}

\bibitem[{Militzer {et~al.}(2022)Militzer, Hubbard, Wahl, Lunine, Galanti, Kaspi, Miguel, Guillot, Moore, Parisi, Connerney, Helled, Cao, Mankovich, Stevenson, Park, Wong, Atreya, Anderson, \& Bolton}]{Militzer2022}
Militzer, B., Hubbard, W.~B., Wahl, S., {et~al.} 2022, The Planetary Science Journal, 3, 185, \dodoi{10.3847/PSJ/ac7ec8}

\bibitem[{Moore {et~al.}(2019)Moore, Cao, Bloxham, Stevenson, Connerney, \& Bolton}]{moore2019}
Moore, K., Cao, H., Bloxham, J., {et~al.} 2019, Nature Astronomy, 3, 730, \dodoi{10.1038/s41550-019-0772-5}

\bibitem[{{Park} {et~al.}(2020){Park}, {Konopliv}, {Ermakov}, {Castillo-Rogez}, {Fu}, {Hughson}, {Prettyman}, {Raymond}, {Scully}, {Sizemore}, {Sori}, {Vaughan}, {Mitri}, {Schmidt}, \& {Russell}}]{Park2020}
{Park}, R.~S., {Konopliv}, A.~S., {Ermakov}, A.~I., {et~al.} 2020, Nature Astronomy, 4, 748, \dodoi{10.1038/s41550-020-1019-1}

\bibitem[{Schneider \& Liu(2009)}]{schneider2009}
Schneider, T., \& Liu, J. 2009, Journal of the atmospheric sciences, 66, 579, \dodoi{10.1175/2008JAS2798.1}

\bibitem[{Showman {et~al.}(2006)Showman, Gierasch, \& Lian}]{Showman2006}
Showman, A.~P., Gierasch, P.~J., \& Lian, Y. 2006, Icarus, 182, 513 , \dodoi{10.1016/j.icarus.2006.01.019}

\bibitem[{Tollefson {et~al.}(2017)Tollefson, Wong, de~Pater, Simon, Orton, Rogers, Atreya, Cosentino, Januszewski, Morales-Juberías, \& Marcus}]{Tollefson2017}
Tollefson, J., Wong, M.~H., de~Pater, I., {et~al.} 2017, Icarus, 296, 163 , \dodoi{10.1016/j.icarus.2017.06.007}

\bibitem[{Tyler(2022)}]{Tyler2022}
Tyler, R.~H. 2022, The Planetary Science Journal, 3, 250, \dodoi{10.3847/PSJ/ac8f91}

\bibitem[{{Vasavada} \& {Showman}(2005)}]{VS2005}
{Vasavada}, A.~R., \& {Showman}, A.~P. 2005, Reports on Progress in Physics, 68, 1935, \dodoi{10.1088/0034-4885/68/8/R06}

\end{thebibliography}
\bibliographystyle{aasjournal}

\end{document}